%% file: main.tex
\documentclass[10pt,twocolumn,letterpaper]{article}

\usepackage{cvpr}      % To produce the REVIEW version
\usepackage{graphicx}
\usepackage{amsmath,amsfonts,amsthm,bm}
\usepackage{amssymb}
\usepackage{booktabs, multirow}
\usepackage{algorithm}
\usepackage{algorithmic}
\usepackage{bbm}
\usepackage{adjustbox}
\usepackage{booktabs}
\usepackage{wrapfig}
\input{math_commands}

\usepackage{subfiles}

\usepackage[pagebackref,breaklinks,colorlinks]{hyperref}

\usepackage[capitalize]{cleveref}
\crefname{section}{Sec.}{Secs.}
\Crefname{section}{Section}{Sections}
\Crefname{table}{Table}{Tables}
\crefname{table}{Tab.}{Tabs.}

\def\cvprPaperID{5578} % *** Enter the CVPR Paper ID here
\def\confName{CVPR}
\def\confYear{2023}

\begin{document}

\title{DiffBP: Generative Diffusion of 3D Molecules for Target Protein Binding}

\author{
Haitao Lin$^{1,*}$\\
\and
Yufei Huang$^{1,*}$\\
\and
Odin Zhang$^1$\\
\and
Siqi Ma$^2$\\
\and
Meng Liu$^3$\\
\and
Lirong Wu$^2$ \\
\and
Xuanjing Li$^4$\\
\and
Shuiwang Ji$^{3}$\\\
\and
Tingjun Hou$^{1,\dagger}$\\
\and
Stan Z. Li$^{2,\dagger}$\\
}
\maketitle
\def\thefootnote{1}\footnotetext{Zhejiang University, Hangzhou, China.}
\def\thefootnote{2}\footnotetext{AI Lab, Research Center for Industries of the Future, Westlake University, Hangzhou, China.}
\def\thefootnote{3}\footnotetext{Texas A\&M University, Texas, U.S.}
\def\thefootnote{4}\footnotetext{Jingdong, Beijing, China.}

\def\thefootnote{*}\footnotetext{These authors contributed equally to this work.}
\def\thefootnote{$\dagger$}\footnotetext{Corresponding authors.}
%%%%%%%%% ABSTRACT
\begin{abstract}
Generating molecules that bind to specific proteins is an important but challenging task in drug discovery. Most previous works typically generate atoms autoregressively, with element types and 3D coordinates of atoms generated one by one. However, in real-world molecular systems, interactions among atoms are global, spanning the entire molecule, leading to pair-coupled energy function among atoms. With such energy-based consideration, modeling probability should rely on joint distributions rather than sequential conditional ones. Thus, the unnatural sequential auto-regressive approach to molecule generation is prone to violating physical rules, yielding molecules with unfavorable properties. In this study, we propose DiffBP, a generative diffusion model that generates molecular 3D structures, leveraging target proteins as contextual constraints at the full-atom level in a non-autoregressive way. Given a designated 3D protein binding site, our model learns to denoise both element types and 3D coordinates of an entire molecule using an equivariant network. In experimental evaluations, DiffBP demonstrates competitive performance against existing methods, generating molecules with high protein affinity, appropriate molecule sizes, and favorable drug-like profiles. Additionally, we developed a website server for medicinal chemists interested in exploring the art of molecular generation, which is accessible at http://www.manimer.com/moleculeformation/index. 
\end{abstract}

%%%%%%%%% BODY TEXT
\section{Introduction}
\label{sec:intro}
Deep learning is revolutionizing the fields like biology \cite{Dauparas2022pro,gao2022alphadesign, YaotingSun2022ArtificialID} and molecular science \cite{xu2021molecule3d, Zeng2022deep}.
In micro-molecule design, a number of works on generating the chemical formulas \cite{shi2019graphaf, polykovskiy2020molecular,tan2022target,yang2017chemts,liu2021graphebm,pmlr-v119-jin20a,zang2020moflow} or conformations of molecules have emerged \cite{townshend2021atom3d, xu2021molecule3d, molsurvey}. 
In macro-molecule design, AlphaFold together with other protein structure prediction methods has had a tremendous and long-standing impact on computational biochemistry \cite{AlphaFold-Multimer2021,AlphaFold2021,rosettafold1,rosettafold2}. 

% Delected: On one hand, classical methods on SBDD are computation-intensive and time-consuming. By contrast, data-driven methods such as Deep Learning can speed up SBDD in dealing with massive data and is just suitable for SBDD.
Success in reverse pharmacology proved that structure-based drug design (SBDD) is a promising method for faster and more cost-efficient lead compound discovery \cite{kuntz1992structure,drews2000drug}.
It directs the discovery of a lead compound which could show potent target inhibitory activity with knowledge about the disease at the molecular level. 
However, it is underexplored to use machine learning approaches to design molecules that can bind to a specific protein as the target. On one hand, massive data is required to develop machine learning approaches. Such datasets are recently available~\cite{francoeur2020three}.
On the other hand, this task is considerably challenging in machine learning for the following three reasons. 
(i) The protein binding site as the conditional context is complicated. Not only do the 3D geometric structures of the target proteins influence the structure and property of the binding molecules, but also other informative contexts should be considered for generating molecules with high affinities, such as the amino acid types.
(ii) The desired distribution over molecular chemistry and coordinates has enormous support sets. Unlike the conformation generation task, the chemical formulas as 2D graph constraints are not known, so the sophisticated coupling of element types, continuous 3D coordinates, and other chemical properties or geometries should be captured by a well-designed model. (iii) Geometric symmetries of molecules should be 
considered for generalization. In the physical 3D space, these symmetries include translations and rotations from the Euclidean group. In particular, if we perform these symmetry operations on a binding site, the generated molecules are expected to be rotated or translated in the same way.  

Recently, a line of auto-regressive methods are proposed for SBDD \cite{masuda2020generating,luo20213d,peng2022pocket2mol,liu2022graphbp}. 
They generate atoms' coordinates and element types one by one, in which the distributions of the next atom's position and type are determined by the former generated ones and the given protein context.
However, several problems exist in such an auto-regressive schema. 
(i) The sequential modeling of a molecule violates real-world physical scenarios, in which the interactions among atoms are global. In other words, each atom's position and element type are affected by all the other atoms in molecular systems.
(ii) The auto-regressive sampling for a molecule usually suffers from \emph{`early stopping'} problem. To be specific, the model tends to generate molecules with a small number of atoms (also called molecule size), thereby failing to capture the true distribution of atom numbers in drug-like molecules.

To tackle these significant limitations, we propose the target-aware molecular Diffusion model for Protein Binding (DiffBP).
% In DiffBP, it employs an equivariant graph neural network to jointly denoise the element types and 3D coordinates of the full atoms in a binding site in the reverse generating process.
By harnessing diffusion denoising generative models' ability to generate samples of high-quality \cite{ho2020denoising,song2020score,cao2022survey}, and great expressivity of equivariant graph neural networks \cite{victor2022egnn,jing2021gvp2}, our model can generate molecules of good drug properties, with high affinity when binding to the target protein.
We summarize our main contributions as:
\begin{itemize}
  \item We first analyze the problems emerging in auto-regressive models from a physics and probabilistic perspective. Motivated by these observations, we propose DiffBP, which directly models the full atoms in target-aware molecule generation.
  \item Delicately-designed procedures are established, considering molecules' center of mass prediction, atom number distribution modeling and other aspects.
  \item Experiments demonstrate that our DiffBP achieves promising performance compared with previous methods, in terms of appropriate molecule size, higher affinity with target protein and other drug properties.
\end{itemize}
%-------------------------------------------------------------------------
\section{Background}
\subsection{Problem Statement}
For a protein-molecule (also called protein-ligand) binding system as $\mathcal{C}$, which contains $N + M$ atoms, we represent the index set of molecules as $\mathcal{I}_{\mathrm{mol}}$ and proteins as $\mathcal{I}_{\mathrm{pro}}$, where $|\mathcal{I}_{\mathrm{mol}}| = N$ and $|\mathcal{I}_{\mathrm{pro}}| = M$.
To be specific, let $\bm{a}_i$ be the $K$-dimensional one-hot vector indicating the atom element type of the $i$-th atom in the binding system and $\bm{x}_i$ be its 3D Cartesian coordinate, and then $\mathcal{C} = \{(\bm{a}_i, \bm{x}_i)\}_{i=1}^{N+M}$  can be split into two sets as $\mathcal{C} = \mathcal{M} \cap \mathcal{P}$, where
$\mathcal{M} = \{(\bm{a}_i, \bm{x}_i): i\in \mathcal{I}_\mathrm{mol}\}$ and $\mathcal{P} = \{(\bm{a}_j, \bm{x}_j): j\in \mathcal{I}_\mathrm{pro}\}$.
For protein-aware molecule generation, our goal is to establish a probabilistic model to learn the conditional distribution of molecules conditioned on the target proteins, \textit{i.e.} $p(\mathcal{M}|\mathcal{P})$.
\paragraph{Problems in auto-regressive models.} 
Recently proposed auto-regressive models sequentially generate $(\bm{a}_i, \bm{x}_i)$ by modeling the conditional probability $p(\bm{a}_i, \bm{x}_i|\mathcal{C}_{i-1})$, where $\mathcal{C}_{i-1} = \{(\bm{a}_j, \bm{x}_j)\}_{j=1}^{i-1} \cup \mathcal{P}$ is the intermediate binding system at the step $i$. By this means, the desired probability is modeled as a sequence of conditional distributions, as $p(\mathcal{M}|\mathcal{P}) = \prod_{i=1}^{N} p(\bm{a}_i, \bm{x}_i|\mathcal{C}_{i-1})$, where $\mathcal{C}_{0} = \mathcal{P}$.  By contrast, in real-world protein-molecule systems, there are force interactions between any pair (or even higher order) of atoms such that the energy function can be decomposed as
    $E(\mathcal{C}) = \sum_{i\neq j} E(\bm{a}_i, \bm{x}_i, \bm{a}_j, \bm{x}_j)$.
The stable system reaches an energy-minimal state. From the perspective of energy-based generative models, the corresponding Boltzmann distribution is written as $p(\mathcal{C}) = \exp{(-E(\mathcal{C})/\kappa\tau)} = \prod_{i\neq j} p(\bm{a}_i, \bm{x}_i, \bm{a}_j, \bm{x}_j)$, where $\kappa$ is Boltzmann constant and $\tau$ is the temperature, indicating that the modeling of probability is based on joint distributions at a full atom level, instead of sequential conditional distributions. Thus, auto-regressive models are likely to violate physical rules. To address it, we employ the following diffusion models for molecule generation at a full-atom level. 
\subsection{Diffusion Models}
\paragraph{Diffusion on continuous variables.}
Diffusion models \cite{sohl2015deep, xu2021geodiff} for continuous variables learn the data distribution by manually constructing the forward diffusion process and using a denoising model to gradually remove the noise added in the diffusion process. The latter process is called the reverse denoising process. 
Denote the input data point by $\bm{z} = \bm{z}_0$, and the diffusion process adds multivariate Gaussian noise to $\bm{z}_t$ for $t = 0,\ldots,T$, so that 
\begin{equation}
 q(\bm{z}_t | \bm{z}_{0}) = \mathcal{N}(\alpha_t\bm{z}_0, \sigma^2_t\bm{I}) , \label{eq:diffbase}
\end{equation} 
where $\alpha_t \in \mathbb{R}^+$ is usually monotonically decreasing from 1 to 0, and $\sigma^2$ is increasing, which means the retained input signals are gradually corrupted by the gaussian noise along $t$-axis, leading to $q(\bm{z}_T) = \mathcal{N}(\bm{0},\bm{I})$.
For variance-preserving diffusion process \cite{ho2020denoising}, $\alpha_t = \sqrt{1-\sigma_t^2}$; For variance-exploding process \cite{song2019generative,song2020score}, $\alpha_t = 1$.
Following recently proposed Variational Diffusion Models \cite{Dieerik2021vdm, emiel2021edm}, where signal-to-noise ratio is defined as 
\begin{equation}
  \mathrm{SNR}(t) = \alpha_t^2/\sigma_t^2,
\end{equation} 
The Markov representation of the diffusion process can be equivalently written with transition distribution as 
\begin{equation}
  q(\bm{z}_t|\bm{z}_s) = \mathcal{N}(\alpha_{t|s} \bm{z}_s, \sigma^2_{t|s}\bm{I}),
\end{equation}
where $s = t-1$, $\alpha_{t|s} = \alpha_t/\alpha_s$ and $\sigma^2_{t|s} = \sigma^2_t - \alpha^2_{t|s}\sigma^2_s$.
In the true denoising process, when the input signals are given, the transition distribution is also normal, and given by 
\begin{equation}
  q(\bm{z}_s|\bm{z}_0, \bm{z}_t) = \mathcal{N}(\bm{\mu}(\bm{z}_0,\bm{z}_t), \sigma'^2_{t|s}\bm{I}) \label{eq:truetrans}
\end{equation}
where $\bm{\mu}(\bm{z}_0,\bm{z}_t)=\frac{\alpha_{t|s}\sigma_s^2}{\sigma_t^2} \bm{z}_t + \frac{\alpha_s \sigma^2_{t|s}}{\sigma_t^2}\bm{z}_0, \sigma'_{t|s} = \frac{\sigma_{t|s}\sigma_s}{\sigma_t}$.

In the generative denoising process, because the input signals are not given, it firstly uses a neural network to approximate $\bm{z}_0$ by $\bm{\hat{z}}_0$,
and then the learned transition distribution which is similar to the Eq.~(\ref{eq:truetrans}) is written as 
\begin{equation}
  p(\bm{z}_s|\bm{z}_t) = \mathcal{N}(\bm{\mu}(\bm{\hat{z}}_0,\bm{z}_t), \sigma'^2_{t|s}\bm{I}). \label{eq:apprtrans}
\end{equation}
Besides, by rewriting the variational lower bound on the likelihood of $\bm{z}_0$, the loss function can be simplified to
\begin{equation}
  L_{\mathrm{cont}} = \sum_{t=1}^{T}\mathrm{KL}(q(\bm{z}_{s}| \bm{z}_0, \bm{z}_t)|p(\bm{z}_s|\bm{z}_t)).
\end{equation}
Instead of directly predicting $\bm{\hat{z}}_0$ using neural networks, diffusion models try to predict the added noise in a score-matching way. Specifically,
according to Eq.~(\ref{eq:diffbase}), $\bm{z}_t = \alpha_t \bm{z}_0 + \sigma_t \bm{\epsilon}_t$, where $\bm{\epsilon}_t\sim \mathcal{N}(\bm{0}, \bm{I})$, and then the neural network $\phi$ predicts $\bm{\hat\epsilon} = \phi(\bm{z}_t, t)$, so that
$\bm{\hat z}_0 = \frac{1}{\alpha_t}\bm{z}_t - \frac{\sigma_t}{\alpha_t}\bm{\hat\epsilon}_t$. Further, the simplified loss function can is written as 
\begin{equation}
  L_{\mathrm{cont}} = \sum_{t=1}^{T}\mathbb{E}_{\bm{\epsilon}_t\sim \mathcal{N}(\bm{0}, \bm{I})}[\frac{1}{2} (1-\frac{\mathrm{SNR}(s)}{\mathrm{SNR}(t)})\lVert \bm{\epsilon}_t - \bm{\hat \epsilon}_t\rVert^2]. \label{eq:loss_cont}
\end{equation}

\paragraph{Diffusion on discrete variables.}
Differing from Gaussian diffusion processes that operate in continuous spaces, diffusion models for discrete variables \cite{jacob2021d3pm,bond2021unleashing} firstly define the diffusion process of random variables ${z}_t \in\{1,\ldots,K\}$ with $K$ categories as 
\begin{equation}
  q(z_t|\bm{z}_0) = \mathrm{Cat}(\bm{z}_0 \bm{\bar Q}_t),
\end{equation}
where $\bm{z}_t$ is the one-hot vector of $z_t$, $\bm{\bar Q}_t = \bm{Q}_1\bm{Q}_2 \ldots \bm{Q}_t$ with $[\bm{Q}_t]_{ij} = q(z_{t}=j|z_{s}=i)$ denoting diffusion transition probabilities, and $[\bm{z}_0 \bm{\bar Q}_t]_i$ is the probability of ${z}_t=i$.
The true denoising process is given as 
\begin{equation}
  q(z_s|\bm{z}_t, \bm{z}_0) = \mathrm{Cat}(\frac{\bm{z}_t\bm{Q}_t^\intercal \odot \bm{z}_0\bm{\bar Q}_{t-1}^\intercal}{\bm{z}_0 \bm{\bar Q}_t \bm{z}_t^\intercal}).
\end{equation} 
Here we employ the absorbing discrete diffusion model, which parameterizes the transition matrix $\bm{Q}_t$ as 
\begin{align}
  \left[\bm{Q}_t\right]_{ij} = \begin{cases}
  1  \quad&\text{if}  \quad i = j = K+1 \\
  1 - \beta_t \quad &\text{if} \quad i = j \ne K+1 \\
  \beta_t \quad &\text{if} \quad j = K+1, i \ne K+1\\
\end{cases},
  \label{eq:cat_forward_kernel_mat_mask}
\end{align}
where $K+1$ is an absorbing state, usually denoted as [MASK] token in text generation. $\beta_t$ monotonically increases from 0 to 1, means that when $t=T$, all the discrete variables are absorbed into the $K+1$ category. 

Due to the effectiveness of BERT-style training, a neural network is used to directly predict $p(z_0|\bm{z}_t)$ rather than $p(z_s|\bm{z}_t)$, leading to a BERT-like training objective as
\begin{equation}
  L_{\mathrm{disc}} = \sum_{t=1}^{T}\lambda(t)\mathbb{E}_{z_t\sim q(z_t|\bm{z}_0)}\left[\log p(z_0|\bm{z}_t)\right],\label{eq:loss_disc}
\end{equation}
where $\lambda(t)$ is the weights at different time step. For example, $\lambda(t) = 1/T$ leads to equal weights. 
\subsection{Equivariance}
For molecular systems, the atoms' positions are represented as 3D Cartesian coordinates, so both \emph{equivariance} of the transition distributions and \emph{invariance} of the data distribution of 3D structure are required for generalization \textit{w.r.t.} the SE(3) group.
Formally, $f: \mathbb{R}^3 \rightarrow \mathbb{R}^3$ is an equivariant function \textit{w.r.t.} SE(3) group, if for any rotation and translation transformation in the group, which is represented by  $\bm{\mathrm{R}}$ as orthogonal matrices and $\bm{\mathrm{t}} \in\mathbb{R}^3$   respectively, $f(\bm{\mathrm{R}} \bm{x} + \bm{\mathrm{t}}) = \bm{\mathrm{R}} f(\bm{x}) + \bm{\mathrm{t}}$.
If $f:\mathbb{R}^3 \rightarrow \mathbb{D}$ is invariant \textit{w.r.t.} SE(3) group, then $f(\bm{\mathrm{R}} \bm{x} + \bm{\mathrm{t}}) = f(\bm{x})$, where $\mathbb{D}$ can be any domain.  

In the setting of generative models, the learned distribution $p(\bm{x})$ should be invariant to SE(3) group \cite{emiel2021edm}.
Kohler \etal \cite{kohler2020ef} showed that an invariant distribution composed with an equivariant invertible function results in an invariant distribution.
In the setting of diffusion models, the transition distribution is defined to be SE(3)-equivariant if $p(\bm{x}_s|\bm{x}_t) = p(\bm{\mathrm{R}}\bm{x}_s + \bm{\mathrm{t}}| \bm{\mathrm{R}}\bm{x}_t + \bm{\mathrm{t}})$.
Xu \etal \cite{xu2021geodiff} showed if $p(\bm{x}_T)$ is an SE(3)-invariant distribution, and for any $t$, $p(\bm{x}_s|\bm{x}_t)$ is equivariant, then $p(\bm{x}_0)$ is also SE(3)-invariant.    
However, because the translation equivariance cannot be preserved in the diffusion process (Appendix.~\ref{app:zeromean}), we fix the center of mass (CoM) to zero to avoid all the translation transformation (details in Sec.~\ref{sec:pregenerate}), and diffuse and denoise the atoms' coordinates in the linear subspace with $\sum_{i\in \mathcal{I}_{\mathrm{mol}}}\bm{x}_i = 0$.
Further, in the reverse process, because the initial distribution of $p(\bm{x}_T)$ is set as standard Gaussian, the distribution naturally satisfies invariance to rotation transformation. The zero CoM trick in the corresponding denoising process also circumvents that $p(\bm{x}_T + \bm{t}) = p(\bm{x}_T)$, which makes it impossible for $p$ to be a distribution.

\section{Method}
\begin{figure*}[]
    \centering
    \includegraphics[width=1.42\linewidth, trim = 53 2030 920 120,clip]{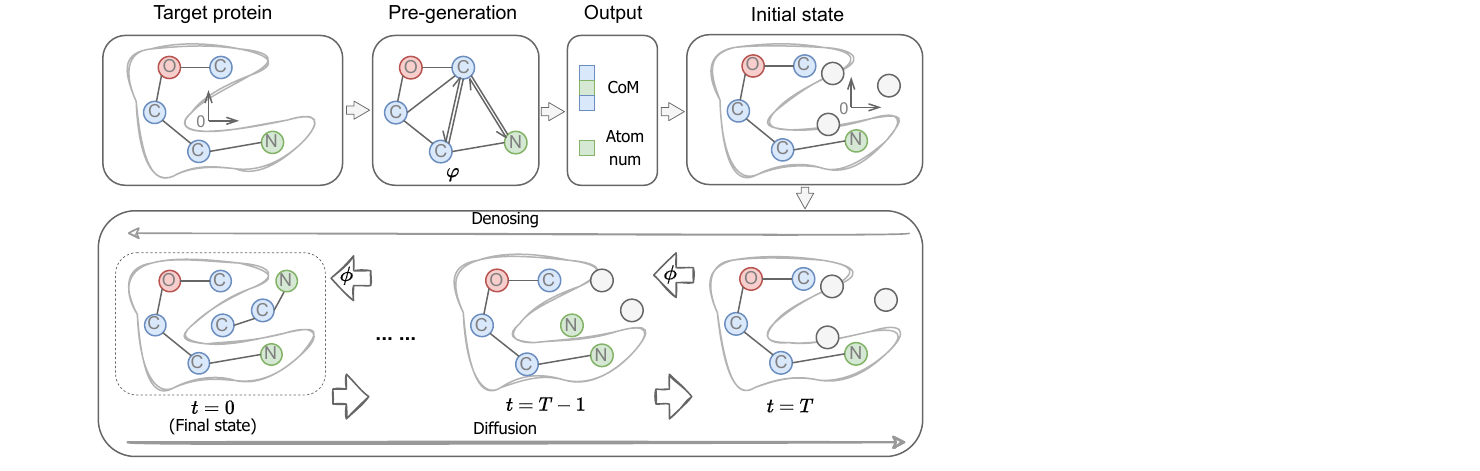}\vspace{-1em}
    \caption{Overall framework as an illustration of the workflows of DiffBP.}\vspace{-1em}
    \label{fig:diffbp_frame}
\end{figure*}
\subsection{Target-aware Diffusion Process}
By using the notations of diffusion models, we write the input binding system as $\mathcal{C}_{0}$, and the intermediate noisy system as $\mathcal{C}_{t}$, where $t=1,\ldots,T$.
Note that both the diffusion and denoising processes are all performed on $\mathcal{M}$, while the protein system $\mathcal{P}$ keeps unchanged in the processes. Therefore, for notation simplicity, we omit the subscript $i\in \mathcal{I}_{\mathrm{mol}}$ in the following and assume that the forward diffusion process as 
\begin{equation}
  q(\bm{x}_{i,t}, \bm{a}_{i,t} | \mathcal{C}_{0}) = q(\bm{x}_{i,t}| \bm{x}_{i,0}, \mathcal{P})q(\bm{a}_{i,t}|\bm{a}_{i,0}, \mathcal{P}).
\end{equation}

\paragraph{Diffusion on continuous positions.}
For the continuous 3D coordinate $\bm{x}_i$ of each atom in the molecule, the forward diffusion process is written as 
\begin{equation}
  \bm{x}_{i,t} = \alpha_t  \bm{x}_{i,0} + \sigma_t \bm{\epsilon}_{i,t},
\end{equation}
where $\bm{\epsilon}_{i,t} \sim \mathcal{N}(\bm{0}, \bm{I})$, where $\bm{0}\in \mathbb{R}^3, \bm{I}\in \mathbb{R}^{3\times 3}$. 
The noise schedule of $\{\alpha_t\}_{t=1}^{T}$ and $\{\sigma_t\}_{t=1}^{T}$ are chosen as a simple polynomial scheme (Appendix.~\ref{app:diffusion_detail}).
To fix the CoM of all the intermediate molecules to zero, we translate $\{\bm{x}_{i,t}\}_{i=1}^{N}$ so that $\sum_{i}\bm{x}_{i,t} = \bm{0}$ for $t=1,\ldots,T$.
\paragraph{Diffusion on discrete types.}
For the discrete atom element type, we use the absorbing diffusion model, where the noise schedule is chosen as uniform. In detail, assume $K+1$ is the absorbing state, and the atom element type $a_{i,t} \in \{1,\ldots,K+1\}$, corresponding to its one-hot vector $\bm{a}_{i,t} \in \{0,1\}^{K+1}$, then
\begin{equation}
  \begin{aligned}
    q({a}_{i,t} &= {a}_{0,t} | \bm{a}_{0,t}) = 1 - \frac{t}{T}; \\
    q({a}_{i,t} &= K+1 | \bm{a}_{0,t}) = \frac{t}{T}.
  \end{aligned}
\end{equation}
\subsection{Equivariant Graph Denoiser}
To learn the transition distribution $p(\bm{x}_{i,t-1}, \bm{a}_{i,t-1}|\mathcal{C}_{t})$, we use the EGNN \cite{victor2022egnn} satisfying rotational-equivariance \textit{w.r.t.} $\bm{x}_i$ and invariance \textit{w.r.t.} $\bm{a}_i$. Specifically,
\begin{equation}
  \begin{aligned}
  &p(a_{i,t-1}, \bm{x}_{i,t-1}|\mathcal{C}_{t}) \\
  =\enspace& p(a_{i,t-1}, \bm{\mathrm{R}}\bm{x}_{i,t-1}|\{( \bm{a}_{j,t},\bm{\mathrm{R}} \bm{x}_{j,t})\}_{j=1}^{N+M}).
  \end{aligned}
\end{equation}
Then a commonly-used SE(3)-EGNN reads 
\begin{equation}
  \begin{aligned}
     \left [\bm{\hat \epsilon}_{i,t},  \bm{\hat p}_{i,0} \right ] = \phi_{\theta}(\{(\bm{a}_{j,t}, \bm{x}_{j,t})\}_{j=1}^{N+M}, t),
  \end{aligned}
\end{equation}
where the $l$-th equivariant convolutional layer is defined as 
\begin{equation}
  \begin{aligned}
  \bm{v}_{i,j} &= \psi_{\mathrm{hid}}(\Vert\bm{x}^{(l)}_i - \bm{x}^{(l)}_j \Vert^2, \bm{h}_i^{(l)}, \bm{h}_j^{(l)}, \bm{e}_{i}, \bm{e}_j, \bm{e}_t); \\
 \bm{h}_{i}^{(l+1)} &= \bm{h}_{i}^{(l)}  + \sum_{j\in \mathcal{V}(i)}\bm{v}_{i,j} ;\\
 \bm{u}_{i,j} &= \psi_{\mathrm{eqv}}(\Vert\bm{x}^{(l)}_i - \bm{x}^{(l)}_j \Vert^2, \bm{h}_i^{(l+1)}, \bm{h}_j^{(l+1)}, \bm{e}_{i}, \bm{e}_j, \bm{e}_t); \\
 \bm{x}_i^{(l+1)} &= \bm{x}_i^{(l)} + \sum_{j\in \mathcal{V}(i)}\frac{(\bm{x}^{(l)}_i - \bm{x}^{(l)}_j)}{\Vert\bm{x}^{(l)}_i - \bm{x}^{(l)}_j\Vert^2} \bm{u}_{i,j},
\label{eq:egnn_struct}
\end{aligned}
\end{equation}
in which $i \in \mathcal{I}_{\mathrm{mol}}$, and $j \in \mathcal{I}_{\mathrm{mol}}\cup \mathcal{I}_{\mathrm{pro}}$. 
The input $\bm{e}_i$ is the atom type embedding of $\bm{a}_i$, $\bm{e}_t$ is the time embedding of $t$, and $\mathcal{V}(i)$ is the neighborhood of $i$ established with KNN according to $\mathcal{C}_t$.
$\bm{h}_i^{(l)} \in \mathbb{R}^{D_h}$ is the $i$-th atom's hidden state in the $l$-th layer which is rotational-invariant. Another alternative
 architecture of the graph denoiser is Geometric Vector Perceptron \cite{jing2021gvp1,jing2021gvp2}. In general, we find these two architectures have close performance empirically.
In the final layer, a softmax function following a multi-layer perceptron $f_{\mathrm{mlp}}: \mathbb{R}^{D_h} \rightarrow \mathbb{R}^{K} $ is used to transform the logits $f_{\mathrm{mlp}}(\bm{h}_i^{(L)})$ into atom element type probabilities $\bm{\hat p}_{i,0}$.

\subsection{Optimization Objective}
\label{sec:loss}
\paragraph{Denoising continuous positions.} As we set the final layer's output as $\bm{\hat \epsilon}_{i,t} =\bm{x}_{i,t}^{(L)}$, and $\mathbb{E}[\bm{\epsilon}_{i,t}] = \bm{0}$, we firstly translate $\bm{\hat \epsilon}_{i,t}$, so that $\sum_i \bm{\hat \epsilon}_{i,t} = \bm{0}$, and then employ Eq.~(\ref{eq:loss_cont}) as our loss on atom's continuous position, which reads 
\begin{equation}
  L_{\mathrm{pos}} = \sum_{t=1}^{T}\sum_{i=1}^{N}\mathbb{E}_{\bm{\epsilon}_{i,t}\sim \mathcal{N}(\bm{0}, \bm{I})}[\frac{1}{2} (1-\frac{\mathrm{SNR}(s)}{\mathrm{SNR}(t)})\lVert \bm{\epsilon}_{i,t} - \bm{\hat \epsilon}_{i,t}\rVert^2]. \label{eq:loss_pos}
\end{equation}
\paragraph{Denoising discrete types.} The rotational-invariant probability vector $\bm{\hat p}_{i,0} = \mathrm{softmax}(f_{\mathrm{mlp}}(\bm{h}_{i}))$ gives the distribution of the $i$-th atom's element type. 
We use Eq.~(\ref{eq:loss_disc}) as the training loss, to recover the atoms of the absorbing type $K+1$ back to its true type with the uniform weights of time.

\begin{equation}
  L_{\mathrm{type}} = \frac{1}{T}\sum_{t=1}^{T}\mathbb{E}_{a_{i,t}\sim q(a_{i,t}|\bm{a}_{i,0})}[\sum_{i=1, \atop a_{i,t} = K+1}^{N}  \mathrm{CE}(\bm{a}_{i,0}, \bm{\hat p}_{i,0})], \label{eq:loss_type} 
\end{equation}
where $\mathrm{CE}(\cdot,\cdot)$ is the cross-entropy loss, which is only calculated on the absorbing types.
\paragraph{Avoiding intersection for binding.} In the binding site, chances are that the noisy atom positions $\bm{x}_t$ go across the surface of a protein, leading to the intersection of proteins and molecules.
In the reverse denoising process, we hope to include the inductive bias of non-intersection. To achieve it, we add an intersection loss as a regularization term in protein docking \cite{octavian2021equidock}. 
We turn to previous works on the surface of proteins and point cloud reconstruction \cite{proteinsurface,proteinsurface2}, where the surface of a protein point cloud $\{\bm{x}_j:j\in\mathcal{I}_\mathrm{pro}\}$ is firstly defined as $\{\bm{x} \in \mathbb{R}^3: S(\bm{x}) = \gamma \}$, where $S(\bm{x}) = -\rho \ln(\sum_{j\in \mathcal{I}_\mathrm{pro}}\exp(-\Vert\bm{x} - \bm{x}_j \Vert ^2 /\rho))$. 
In this way, $\{\bm{x} \in \mathbb{R}^3: S(\bm{x}) < \gamma \}$ is the interior of the protein, and the atoms of the binding molecule should be forced to lay in $\{\bm{x} \in \mathbb{R}^3: S(\bm{x}) > \gamma \}$.
As a result, the inductive bias as a regularization loss function reads
\begin{equation}
  L_{\mathrm{reg}} = \sum_{i=1}^N \max(0, \gamma - S(\bm{\hat x}_{i,0})),
\end{equation} 
where $\bm{\hat x}_{i,0}$ is the approximated positions at $t=0$, as $\bm{\hat x}_{i,0} = \frac{1}{\alpha_t}\bm{x}_{i,t} - \frac{\sigma_t}{\alpha_t}\bm{\hat\epsilon}_{i,t}$. $\gamma$ and $\rho$ are predefined, where we give the choice of them in Sec.~\ref{sec:moduleana}.

\paragraph{Reconstruction on other attributes.} In the denoising step of $t=1$, our equivariant graph denoiser can also be used to recover other attributes of the binding systems, which is based on $\bm{x}_{i,0} \approx \bm{x}_{i,1}$ as the $\mathrm{SNR}(1) \rightarrow \infty$.
For example, in order to predict the binary atom attribute $z^{\mathrm{aro}}$ of `\texttt{is$\_$aromatic}', another prediction head $g_{\mathrm{mlp}}: \mathbb{R}^{D_h} \rightarrow \mathbb{R}$ can be defined, such that 
$p(z^{\mathrm{aro}}_i = 1 | \mathcal{C}_{1}) = p_{i}^{\mathrm{aro}} = \mathrm{sigmoid}(g_{\mathrm{mlp}}(\bm{h}_{i,1}^{(l)}))$. And the binary cross-entropy loss can be used to train the denoiser, leading to 
\begin{equation}
  L_{\mathrm{rec}} = \sum_{i=1}^{N}\mathrm{BCE}(p_{i}^{\mathrm{aro}}, y_i^{\mathrm{aro}}),
\end{equation}
where $y_i^{\mathrm{aro}}$ is the $i$-th atom's ground-truth label on the attribute of `\texttt{is$\_$aromatic}'.
\\
\\
To sum up, the overall loss function used in the training process reads 
\begin{equation}
  L = L_{\mathrm{pos}} + L_{\mathrm{type}} + L_{\mathrm{reg}}+L_{\mathrm{rec}}. \label{eq:overall_loss}
\end{equation}
\subsection{Generative Denoising Process}
\paragraph{Generating atom positions.} In generating the atom positions, we firstly sample $\bm{x}_{i,T} \sim \mathcal{N}(\bm{0}, \bm{I})$. 
Then $\bm{x}_{i,s}$ is drawn from $p(\bm{x}_{i,s} | \bm{x}_{i,t}) = \mathcal{N}(\bm{\mu}(\bm{\hat{x}}_{i,0},\bm{x}_{i,t}), \sigma'^2_{t|s}\bm{I})$, where the parameters is calculated by 
\begin{equation}
  \begin{aligned}
  \mu(\bm{\hat x}_{i,0}, \bm{x}_{i,t}) = \enspace& \frac{\alpha_{t|s}\sigma_s^2}{\sigma_t^2} \bm{x}_{i,t} + \frac{\alpha_s \sigma^2_{t|s}}{\sigma_t^2} \bm{\hat x}_{i,0}\\
  =\enspace&  \frac{1}{\alpha_{t|s}}\bm{x}_{i,t} - \frac{\sigma^2_{t|s}}{\alpha_{t|s}\sigma_{t}} \bm{\hat \epsilon}_{i,t};\\
  \sigma'_{t|s} =\enspace& \frac{\sigma_{t|s}\sigma_s}{\sigma_t} \label{eq:generate_pos}.
\end{aligned}
\end{equation}

\paragraph{Generating element types.} For atoms' element types, $\{a_{i,T}\}$ are all set as the absorbing type of $K+1$ at time $T$. Then, in each step at $t$, we randomly select $(T-t)/T$ of atoms to predict their element types.
Besides, if the element type of an atom has been recovered in the past steps, it would not change, which indicates that the recovery of the element type of each atom is only performed once.  
\begin{algorithm}[t]
  \caption{Training DiffBP \small{(Sampling in Appendix.~\ref{app:diffusion_detail})}}
  \label{alg:optimize_edm}
\begin{algorithmic}
\STATE {\bfseries Input:} Zero-centered molecule $\{\bm{a}_j,\bm{x}_j\}_{j=1}^{N}$, protein $\{\bm{a}_j,\bm{x}_j\}_{j=1}^{M}$, and graph denoiser $\phi_{\theta}$
\STATE Sample $t \sim \mathcal{U}(0, \ldots, T)$, $\bm{\epsilon}_{i,t} \sim \mathcal{N}(\bm{0}, \bm{I})$
\STATE Subtract center of mass from $\bm{\epsilon}_{i,t}$
\STATE Compute $\bm{x}_{i,t} = \alpha_t \bm{x}_{i,0} + \sigma_t \bm{\epsilon}_{i,t}$
\STATE Sample uniformly $\mathcal{I}_{\mathrm{abs}}$, $|\mathcal{I}_{\mathrm{abs}}|/M = t/T$
\IF {$i\in \mathcal{I}_{\mathrm{abs}}\cap\mathcal{I}_{\mathrm{mol}}$}
\STATE Set $a_{i,t} = K + 1$
\ENDIF
\STATE Compute $[\bm{\hat \epsilon}_{i,t},  \bm{\hat p}_{i,0}] = \phi_{\theta}(\{(\bm{a}_{j,t}, \bm{x}_{j,t})\}_{j=1}^{N+M}, t)$
\STATE Minimize $L$ as defined in Eq.~(\ref{eq:overall_loss})
\end{algorithmic}
Note that $\mathcal{I}_{\mathrm{abs}}$ is the index set of atoms of absorbing type.
\end{algorithm}

\paragraph{Pre-generation models.}
\label{sec:pregenerate}
It is noted that the translational invariance cannot be satisfied without the zero CoM trick.
Moreover, there are several mass centers of the system, such as the protein's CoM and the protein-molecule CoM.
Choosing a proper CoM of the system is a matter of careful design. We propose that for the diffusion model, the zero CoM should be performed according to the molecule.
As shown in Eq.~(\ref{eq:generate_pos}),  it can be derived that $\mathbb{E}[\bm{x}_{i,s}] = \frac{1}{\alpha_{t|s}}\mathbb{E}[\bm{x}_{i,t}] - \frac{\sigma^2_{t|s}}{\alpha_{t|s}\sigma_{t}} \mathbb{E}[\bm{\hat \epsilon}_{i,t}]$.
For a well-trained equivariant graph neural network, it is expected that $\mathbb{E}[\Vert\bm{\hat \epsilon}_{i,t} - \bm{\epsilon}_{i,t}\Vert ^2] \rightarrow \bm{0}$.
By Jenson's inequality, $(\mathbb{E}[\Vert\bm{\hat \epsilon}_{i,t} - \bm{\epsilon}_{i,t}\Vert])^2  \leq \mathbb{E}[\Vert\bm{\hat \epsilon}_{i,t} - \bm{\epsilon}_{i,t}\Vert ^2]$,
which indicates that $\mathbb{E}[\bm{\hat \epsilon}_{i,t}] \approx \bm{0}$ for any $i \in \mathcal{I}_{\mathrm{mol}}$ and $0\leq t \leq T$.
Combining with $\mathbb{E}[\bm{x}_{i,T}] = \bm{0}$, we can easily obtain that $\mathbb{E}[\bm{x}_{i,t}] \approx \bm{0}$. 
Therefore, we choose the mass center of the binding molecules as zero by translating the protein-molecule binding system such that the origin of the global coordinate system coincides with the molecule's CoM. Otherwise, it is challenging for the generated molecules to be located in the binding site.

Another problem raised by this translation is that the CoM of the molecules is unknown in the generation process, since only the information of proteins is the pre-given context as conditions.
Besides, as a full-atom generation method, it is necessary to assign the atom numbers of the binding molecules before both diffusion and denoising process.
To address these problems, an additional graph neural network $\varphi_\omega(\mathcal{P}) \in \mathbb{R}^3 \times \mathbb{Z}^{+}$ with the same intermediate architectures as shown in Eq.~(\ref{eq:egnn_struct}) is pre-trained as a pre-generation model, which aims to generate the atom numbers as well as the molecules' CoM before the denoising process. 
In detail,
\begin{equation}
  [\bm{x}_{\mathrm{c}}^\mathrm{mol}, N^\mathrm{mol}] = \varphi_\omega(\{\bm{a}_{j}, \bm{x}_j\}_{j\in \mathcal{I}_{pro}}),
\end{equation}
where $\bm{x}_{\mathrm{c}}^\mathrm{mol}$ is finally obtained by an SE(3)-equivariant pooling as $\mathrm{mean}(\{\bm{x}^{(L)}_j\}_{j=1}^{M}) = \frac{1}{M}\sum_{j=1}^{M}\bm{x}^{(L)}_j $. 
Moreover, an SE(3)-invariant feature $\mathrm{max}(u_{\mathrm{mlp}}(\{\bm{h}^{(L)}_j\}_{j=1}^{M})$ is used to predict the distribution of atom numbers with $u_{\mathrm{mlp}} : \mathbb{R}^{D_h} \rightarrow \mathbb{R}^{N_s}$, where $N_s$ is the number of different molecule sizes which can be statistically obtained by the training set, and $u_{\mathrm{mlp}}$ transforms the latent features into logits for calculating the probability. A dictionary is used to map the predicted class $\hat n_s$ to $\mathbb{Z}^{+}$. For example, $\mathrm{Dict} =\{1:18, 2:27\}$ means when $\hat n_s =1$, the atom number $N^{\mathrm{mol}}$ will be assigned as $18$. 
Once the molecule's CoM and atom number are obtained, we translate the protein system by $\bm{x}_j := \bm{x}_j - \bm{x}_{\mathrm{c}}^\mathrm{mol}$ for $j=1,\ldots,M$, and set $N := N^\mathrm{mol}$ as the molecule size for generation of the diffusion model. The overall workflow of DiffBP is presented in Fig.~\ref{fig:diffbp_frame}.

\section{Related Work}
\paragraph{3D molecule generation.} Machine-learning-based methods have emerged explosively in recent years thanks to development in graph neural networks \cite{gcn,mpnn,liu2021dig,wu2021selfsuper,liu2021spherical} and generative models \cite{doersch2016tutorial,goodfellow2014generative,dinh2016density,kingma2018glow,nielsen2020survae,bengio2021gflownet}. Molecular conformation generation aims to generate 3D structures of molecules with stability, given chemical formulas \cite{xu2021end,shi2021learning,luo2021predicting}. For example, score-based methods including diffusion models \cite{xu2021geodiff, xu2021learning} are generally employed as probabilistic models for conformation generation. Further, De novo molecular generation attempts to generate both 2D chemical formulas and 3D structures from scratch \cite{Victor2021enf,emiel2021edm,luo2021autoregressive}.
\paragraph{Structure-based drug design.}
Increasing scientific interests have been raised in SBDD, which aims to generate both 2D chemical formulas and 3D structures conditioned on target protein structure as contextual constraints. For example, LiGAN \cite{masuda2020generating} and 3DSBDD \cite{luo20213d} are grid-based models which predict whether the grid points are occupied by specific atoms.  Further, Pocket2Mol \cite{peng2022pocket2mol} and GraphBP \cite{liu2022graphbp} generate atoms auto-regressively and directly model the probability of the next atom's type as discrete categorical attribute and position as continuous geometry, achieving success in generating the molecules of high validity and affinity with the target proteins.
\section{Experiments}
\subsection{Setup}
\paragraph{Datasets.} For evaluation of our model, we follow the previous works \cite{masuda2020generating,liu2022graphbp} and use the CrossDocked2020 \cite{francoeur2020three} dataset to generate ligand molecules that bind to target proteins' pockets based on the pocket structures, with the same split of training and test set. For atom-level contexts of protein pockets, we employ element types and the amino acids the pocket atoms belong to as their SE(3)-invariant features, and atoms' positions as SE(3)-equivariant features.  No additional contexts on ligands are used except atom element types and positions in training because in generating process, the only given contexts are pocket atoms' features. 
\paragraph{Methods.} Three state-of-the-art methods including 3DSBDD \cite{luo20213d}, Pocket2Mol \cite{peng2022pocket2mol} and GraphBP \cite{liu2022graphbp} are employed as benchmarks. These three methods are all auto-regressive generative models. Pocket2Mol and GraphBP directly generate the continuous coordinates of atoms. In comparison, 3DSBDD generates the atoms' position on regular grids. We generate 100 molecules for each protein pocket to calculate metrics for further evaluation.

\begin{table*}[]
\caption{Comparison on Binding Affinity of different methods.}
\vspace{-0.5em}
\label{tab:bindcomp}
\resizebox{1.0\linewidth}{!}{
\begin{tabular}{lrrrrrrrrrrrr}
\toprule
            & \multicolumn{3}{c}{3DSBDD}        & \multicolumn{3}{c}{Pocket2Mol}    & \multicolumn{3}{c}{GraphBP}       & \multicolumn{3}{c}{DiffBP}   \\
\midrule
            & \multicolumn{1}{c}{Ratio}   & \multicolumn{1}{c}{MPBG}    & \multicolumn{1}{c}{$\Delta$Binding} & \multicolumn{1}{c}{Ratio}   & \multicolumn{1}{c}{MPBG}    & \multicolumn{1}{c}{$\Delta$Binding} & \multicolumn{1}{c}{Ratio}   & \multicolumn{1}{c}{MPBG}    & \multicolumn{1}{c}{$\Delta$Binding} & \multicolumn{1}{c}{Ratio} & \multicolumn{1}{c}{MPBG} & \multicolumn{1}{c}{$\Delta$Binding} \\
\cmidrule(lr){2-4} \cmidrule(lr){5-7} \cmidrule(lr){8-10} \cmidrule(lr){11-13}
Small  & 41.45\% & 27.92\% & 4.90\%        & 36.62\% & 25.18\% & 4.10\%        & 27.72\% & 35.16\% & 5.19\%       &5.22\%       &17.61\%      & 10.25\%              \\
Medium & 54.06\% & 19.78\% & 14.84\%       & 59.02\% & 5.38\% & 32.53\%       & 32.03\% & 18.68\% & 15.30\%       &75.19\%       &2.36\%      & 40.20\%              \\
Large  & 4.48\%  & -7.53\% & 48.56\%       & 4.36\%  & -11.21\% & 75.42\%       & 37.97\% & -10.13\% & 60.21\%       &19.59\%       &-4.11\%       & 52.64\%              \\
\midrule
Overall     &         & 21.92\%        &12.22\%       &        &11.90\%         &  23.98\%                      &        &12.30\%          & 29.54\%      &       & 1.88\%     &   41.07\%           \\
\bottomrule
\end{tabular}
}\vspace{-0.5em}
\end{table*}

\paragraph{Metrics.} Following GraphBP and LiGAN, for the calculation of Affinity Score, we adopt Gnina \cite{ragoza2017protein, mcnutt2021gnina} which is an ensemble of CNN scoring functions that were trained on the CrossDocked2020 dataset. Empirical studies show such CNN predicted affinity is more accurate than Autodock Vina empirical scoring function. Quantitatively, two metrics on Affinity Score are reported: (i) \textbf{$\bm{\Delta}$Binding} as the percentage of generated molecules that have higher predicted binding affinity than the corresponding reference molecules. It measures the maximum performance that the model can achieve, and the \emph{larger} the \emph{better}.  (ii) Mean Percentage Binding Gap (\textbf{MPBG}) which is proposed in our paper to measure the mean performance of the model. For a single pair of a protein pocket and its reference molecule, it is calculated by $\frac{1}{N_\mathrm{gen}}\sum_{i=1}^{N_\mathrm{gen}}\frac{\mathrm{Aff}^{\mathrm{ref}} - \mathrm{Aff}_i^{\mathrm{gen}}}{\mathrm{Aff}_{\mathrm{ref}}} \times 100\%$. A \emph{smaller} MPBG represents the \emph{better} binding affinity of the generated molecules on average.
Besides, other scores are used for further comparison including (i) \textbf{QED} as quantitative estimation of drug-likeness; (ii) \textbf{SA} as synthetic accessibility score; (iii) \textbf{Sim} as the average Tanimoto similarities of the multiple generated molecules in one pocket, reviling the diversity of generation; (iv) \textbf{LPSK} as the ratio of the generated drug molecules satisfying the Lipinski’s rule of five. These chemical properties are calculated by RDKit package \cite{landrum2006rdkit}. After the molecules are generated by the model, Openbabel\cite{o2011open} is used to construct chemical bonds between atoms and Universal Force Field (UFF) minimization\cite{rappe1992uff} is used for refinement before the metric calculation. Note that in Pocket2Mol, refinement with UFF is not performed, so we report the metric of Pocket2Mol-without-UFF in Appendix.~\ref{app:exp}. And the reported metrics in this part are all based on Pocket2Mol-with-UFF.
\subsection{Property Evaluation}
\begin{figure*}[t]
    \centering
    \includegraphics[width=1.00\linewidth, trim =310 00 310 20,clip]{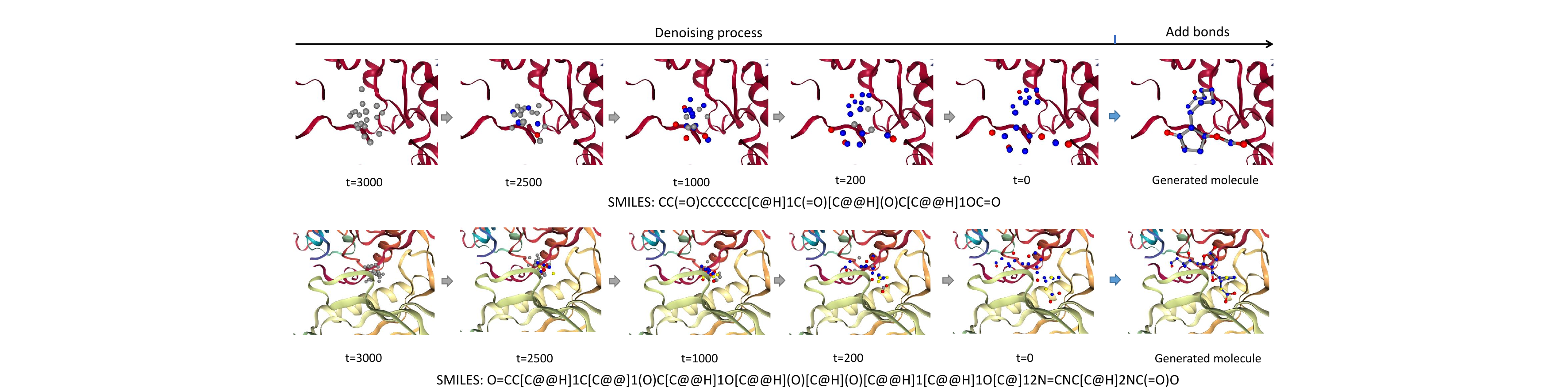}
    \caption{Visualization on generation process two molecules (Affinity score = $4.583$ and $5.682$) which are binding to the protein `1afs$\_$A$\_$rec' and `4azf$\_$A$\_$rec' respectively.}\label{fig:moldynamics}\vspace{-0.5em}
    \label{fig:dynamics1}
\end{figure*}
There are two problems we want to figure out in this part:

\begin{itemize}\vspace{-0.5em}
    \item[(i)] Can DiffBP generate molecules of high binding affinity with target protein at a full-atom level?
\vspace{-0.2em}
    \item[(ii)] Is the distribution of atom number of molecules generated by DiffBP close to drug-like molecules?
\end{itemize}
\vspace{-0.2em}
Guiding by these two questions, we first define the generated molecule sizes at three levels: small, medium, and large, according to the reference molecules' sizes, as 
\begin{equation}
\begin{aligned}
        \mathcal{X}^{\mathrm{Lar}} &= \{\mathcal{\hat M}: |\mathcal{\hat M}| - |\mathcal{M}^{\mathrm{ref}}|>\delta \} \\
        \mathcal{X}^{\mathrm{Med}} &= \{\mathcal{\hat M}: \left\vert|\mathcal{\hat M}| - |\mathcal{M}^{\mathrm{ref}}|\right\vert\leq \delta \}\\
        \mathcal{X}^{\mathrm{Sma}} &= \{\mathcal{\hat M}: |\mathcal{\hat M}| - |\mathcal{M}^{\mathrm{ref}}|<-\delta \}
\end{aligned}
\end{equation}
where $\mathcal{\hat M}$ is the sampled molecules. We set the threshold $\delta$ as 6 in practice. Table.~\ref{tab:bindcomp} presents the two quantitative binding metrics of the four methods, where `Ratio' is the proportion of molecules of different sizes to the total number of molecules generated. Conclusions can be reached as:
\vspace{-0.4em}
\begin{itemize}
    \item 3DSBDD and Pocket2Mol both suffer from `early-stopping' problem of auto-regressive generating process, in which the small ratio is at a high value in comparison with the other two. In contrast, 75.19\% of the molecules generated by DiffBP are of medium size, achieving the lowest MPBG and highest $\Delta$Binding.  
    \vspace{-0.2em}
    \item It is notable that the calculation of binding scores tends to give larger molecules a higher affinity because larger molecules are more likely to make more interactions with the target proteins, which is a common problem of preference of existing scoring functions. \footnote{\url{https://github.com/gnina/gnina/issues/167}}  Therefore, the overall binding score of GraphBP  
   benefits from large proportions of generated molecules with large sizes. However, it also suffers from the distribution shift of atom sizes, \textit{i.e.} the medium ratio is very small while large molecules account for a substantial part of generated samples.
\end{itemize}
\vspace{-0.5em}
\begin{table}[h]\centering
\caption{Other properties of drug-like molecules for comparison. \textbf{Bold values} are the top-2 metrics.  }
\vspace{-0.6em}
\label{tab:drugproper}
\resizebox{0.97\linewidth}{!}{
\begin{tabular}{lrrrr}
\toprule
     & \multicolumn{1}{c}{3DSBDD} & \multicolumn{1}{c}{Pocket2Mol} & \multicolumn{1}{c}{GraphBP} & \multicolumn{1}{c}{DiffBP} \\
\midrule
QED ($\uparrow$)  & 0.3811 & \textbf{0.5106}     & 0.3830  & \textbf{0.4431} \\
SA ($\uparrow$)  & 4.7033 & 5.3125     & \textbf{5.8610}  & \textbf{6.0308} \\
Sim ($\downarrow$) & 0.3485 & 0.4104     &\textbf{0.2707}  & \textbf{0.3290} \\
LPSK ($\uparrow$) & 0.6678 & \textbf{0.8134}     & 0.5961  & \textbf{0.7042} \\
\bottomrule
\end{tabular}
}\vspace{-0.4em}
\end{table}
Although the reference molecules are not the gold standard in the task of SBDD, they can reflect some properties of the binding site, e.g. the site size can be reflected by the reference molecule size to some degree. Therefore, using the reference molecule size to define the appropriate atom numbers of generated molecules is reasonable, and comparing the metrics in the groups of different sizes shows the preference for the molecules' sizes of the scoring functions of drugs' properties.  Other metrics for drug properties on generated molecules are shown in Table.~\ref{tab:drugproper}. It shows that all the metrics are competitive for the molecules generated by DiffBP. However, the scoring functions of different metrics also have their preference for molecule sizes, we give the details in Appendix.~\ref{app:exp}. Fig.~\ref{fig:moldynamics} are two generation examples.

\subsection{Sub-structure Analysis}
We evaluate the ratio of different bond types and molecules containing different rings of the generated molecules and reference, to give a comparison of different methods' generation in terms of sub-structure, shown in Table.~\ref{tab:substructure}. For atom type analysis, see Appendix.~\ref{app:exp}. As presented, DiffBP tends to generate more bonds generally, as well as `Pent' rings, which is an immediate consequence of larger atom numbers of the generated molecules. Pocket2Mol has the advantage to generate fewer `Tri' rings and more `Hex' rings, in coordination with experimental analysis in \cite{peng2022pocket2mol}. 
% \usepackage{multirow}
\begin{table}[h]
\vspace{-0.5em}
\caption{Sub-sturcture analysis on ratios, which is defined as how many sub-structures exist in one molecule on average. \textbf{Bold values} are the top-2 ratios close to reference molecules in training and test sets.  } 
\label{tab:substructure}
\vspace{-0.4em}
\resizebox{1.0\linewidth}{!}{
\begin{tabular}{l|lrrrrrr}
\toprule
                      &        & \multicolumn{1}{c}{Train} & \multicolumn{1}{c}{Test} & \multicolumn{1}{c}{3DSBDD} & \multicolumn{1}{c}{Pocket2Mol}     & \multicolumn{1}{c}{GraphBP}       & \multicolumn{1}{c}{DiffBP}         \\
\midrule
\parbox[t]{2mm}{\multirow{3}{*}{\rotatebox[origin=c]{90}{Bond}}} & Single & 18.76      & 18.24     & \textbf{14.37}  & \textbf{16.50} &                        27.16         & 23.37 \\
                      & Double & 6.67       & 5.24      & 2.42   & \textbf{2.98}  & 0.94          & \textbf{3.43}  \\
                      & Triple & 0.05       & 0.02      & 0.01   & 0.01           & 0.01          & \textbf{0.02}  \\
\midrule
\parbox[t]{2mm}{\multirow{4}{*}{\rotatebox[origin=c]{90}{Ring}}} & Tri    & 0.04       & 0.04      & 1.99   & \textbf{0.83}  & 2.48          & \textbf{0.91}  \\
                      & Quad   & 0.01       & 0.00      & 0.46   & \textbf{0.01}  & 0.93          & \textbf{0.22}  \\
                      & Pent   & 0.8        & 0.71      & 0.31   & 0.43           & \textbf{0.47} & \textbf{0.58}  \\
                      & Hex    & 1.93       & 1.58      & 0.59   & \textbf{1.43}  & 0.38          & \textbf{0.89} \\
\bottomrule
\end{tabular}
}
\vspace{-0.4em}
\end{table}
\subsection{Further Analysis on Modules}
\label{sec:moduleana}
\begin{figure}[h]
    \vspace{-0.4em}
    \centering
    \includegraphics[width=0.93\linewidth, trim =10 00 10 20,clip]{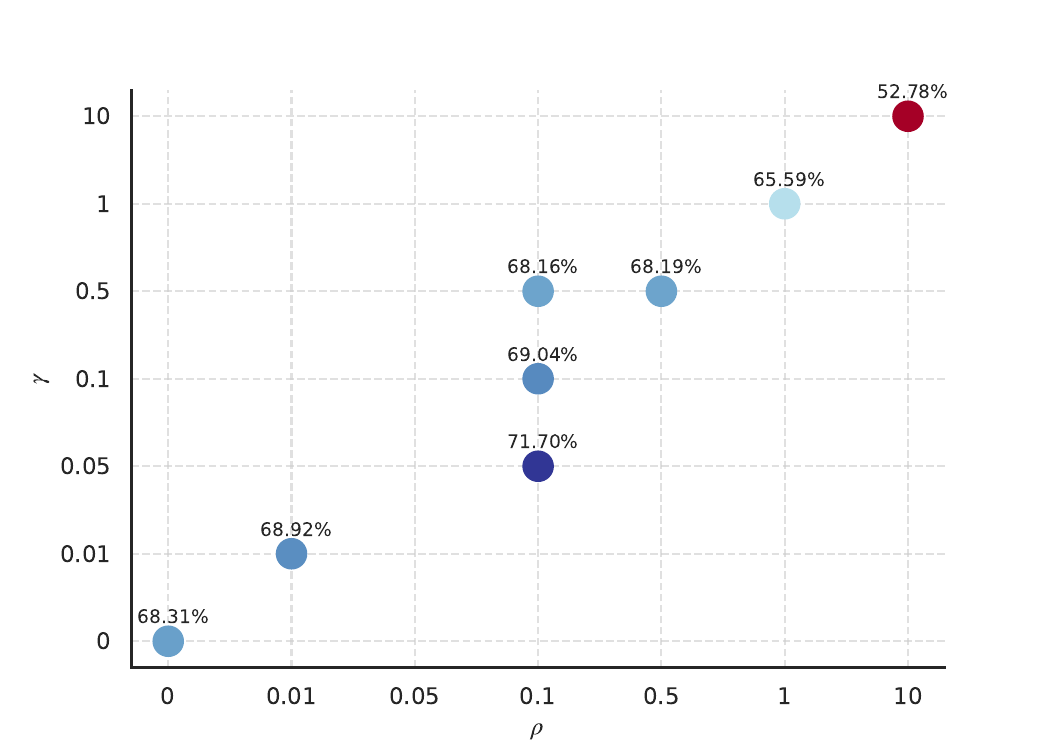}
    \vspace{-0.4em}
    \caption{Validity of generated samples \textit{v.s.} $\rho$ and $\gamma$}
    \label{fig:lossparam}
    \vspace{-0.4em}
\end{figure}

Several modules are not necessary to fill the task in our workflow. Here we attempt to figure out whether they can bring improvements to the performance. The first is the intersection loss as regularization as proposed in Sec.~\ref{sec:loss}. The regularization term affects the success rate (\textit{a.k.a.} validity) greatly. Here we give the validity of the molecules obtained by the trained model \textit{v.s.} different loss parameters, as shown in Figure.~\ref{fig:lossparam}. It shows that the regularization loss can bring improvements to generation validity slightly. However, when the two coefficients are too large, the validity decreases. The rationale for it is that the loss term can be viewed as a repulsive force from protein atoms to molecule atoms, and when the repulsion is too large, the positions of atoms are likely to collapse into a small region, causing the number of each atom's neighbors too large, \textit{i.e.} the valences of atoms are larger than allowed. 

Secondly, the CoMs and atom numbers obtained by pre-generation models in our workflows can be replaced by the ones given by reference molecules, leading to the variant of DiffBP(Pre-Ref). If the generated CoMs stray too far off the binding sites, or the molecule sizes are too large, the contextual information of protein pockets will hardly affect the generative denoising process effectively.  As shown in Table.~\ref{tab:pregen}, our pipeline of DiffBP(Pre-Gen) consisting of a pre-generation model and a diffusion generative model works well since the molecules generated by DiffBP(Pre-Gen) achieve comparable affinity scores to DiffBP(Pre-Ref).
\begin{table}[]\vspace{-0.4em}
\caption{Affinity of DiffBP(Pre-Gen) and DiffBP(Pre-Ref).}
\vspace{-0.5em}
\label{tab:pregen}
\resizebox{1.0\linewidth}{!}{
\begin{tabular}{lrrrrrr}
\toprule
            & \multicolumn{3}{c}{DiffBP(Pre-Gen)}        & \multicolumn{3}{c}{DiffBP(Pre-Ref)}    \\
\midrule
            & \multicolumn{1}{c}{Ratio}   & \multicolumn{1}{c}{MPBG}    & \multicolumn{1}{c}{$\Delta$Binding} & \multicolumn{1}{c}{Ratio}   & \multicolumn{1}{c}{MPBG}    & \multicolumn{1}{c}{$\Delta$Binding} \\
\cmidrule(lr){2-4} \cmidrule(lr){5-7} 
Small  &5.22\%       &17.61\%      & 10.25\%         & 0.00\% & 0.00\% & 0.00\%             \\
Medium &75.19\%      &2.36\%      & 40.20\%        & 100.00\% & 1.56\% & 42.18\%  \\
Large  &19.59\%        &-4.11\%       & 52.64\%       & 0.00\%  & 0.00\% & 0.00\%       \\
\midrule
Overall     &           & 1.88\%     &   41.07\%       &        &-1.56\%        &  42.18\%   \\
\bottomrule
\end{tabular}
}\vspace{-1em}
\end{table}
\section{Conclusion}
A 3D molecule generative model based on diffusion denoising models called DiffBP is proposed, to generate drug-like molecules of high binding affinities with the target proteins. The generation process is in line with the laws of physics, \textit{i.e.} generate molecules at a full atom level. Limitations still exist, including the relatively low validity and deviation from the real drug molecule in the sub-structures.
%%%%%%%%% REFERENCES
{\small
\bibliographystyle{ieee_fullname}
\bibliography{egbib}
}
\subfile{appendix.tex}

\end{document}

%% file: math_commands.tex
\usepackage{amsmath,amsfonts,bm}

\def\eqref#1{equation~\ref{#1}}
\def\1{\bm{1}}

\def\vepsilon{{\bm{\epsilon}}}

\DeclareMathAlphabet{\mathsfit}{\encodingdefault}{\sfdefault}{m}{sl}
\SetMathAlphabet{\mathsfit}{bold}{\encodingdefault}{\sfdefault}{bx}{n}

%% file: appendix.tex
\renewcommand\thefigure{A\arabic{figure}}
\renewcommand\thetable{A\arabic{table}}
\setcounter{table}{0}
\setcounter{figure}{0}
\clearpage
\appendix

\section{Methods}
\subsection{Analysis on invariance in diffusion models}
In the forward diffusion process for positions, which reads 
\begin{equation}
    \bm{x}_{i,t} = \alpha_t \bm{x}_{i,0} + \sigma_t\bm{\epsilon}_{i,t}
\end{equation}
If a translation $\bm{t}$ is performed on the input positions and the added noise, then
\begin{equation}
    \alpha_t (\bm{x}_{i,0} + \bm{\mathrm{t}}) + \sigma_t(\bm{\epsilon}_{i,t}+ \bm{\mathrm{t}}) = (\bm{x}_{i,t} + \bm{\mathrm{t}}) + (1-\alpha_t) \bm{\mathrm{t}} + \sigma_t \bm{\mathrm{t}}.
\end{equation}
 It indicates that the intermediate diffusion states are not translational-equivariant. Then, the translation equivariance in the reserve process is not meaningful. For this reason, we fix the CoM of the input positions and each intermediate state, so $\bm{\mathrm{t}}$ will always equal $0$. The zero-CoM trick circumvents the problem.
 
By contrast, for a rotation transformation, 
\begin{equation}
    \alpha_t (\bm{\mathrm{R}}\bm{x}_{i,0}) + \sigma_t(\bm{\mathrm{R}}\bm{\epsilon}_{i,t}) = \bm{\mathrm{R}}\bm{x}_{i,t}.
\end{equation}
As $\bm{\epsilon}_{i,t} \sim \mathcal{N}(\bm{0}, \bm{1})$, by this way 
\begin{equation}
    p(\bm{\mathrm{R}}\bm{\epsilon}_{i,t}) = p(\bm{\epsilon}_{i,t}),
\end{equation}
and thus 
\begin{equation}
q(\bm{\mathrm{R}}\bm{x}_{i,t}|\bm{\mathrm{R}}\bm{x}_{i,0}) = \mathcal{N}(\alpha_t(\bm{\mathrm{R}}\bm{x}_{i,0}),\sigma_t\bm{I} )
\end{equation}
For the true position distribution, $q(\bm{\mathrm{R}}\bm{x}_{i,0}) = q(\bm{x}_{i,0})$
\begin{equation}
\begin{aligned}
q(\bm{\mathrm{R}}\bm{x}_{i,t}) =& q(\bm{\mathrm{R}}\bm{x}_{i,t}|\bm{\mathrm{R}}\bm{x}_{i,0})q(\bm{\mathrm{R}}\bm{x}_{i,0}) \\ 
=& q(\bm{x}_{i,t})      
\end{aligned}
\end{equation}
In this way, the intermediate states diffusion process satisfies the rotation invariance. And to learn the reverse process, we only require $p(\bm{\mathrm{R}}\bm{x}_{i,0}|\bm{\mathrm{R}}\bm{x}_{i,t})$ to be rotational-equivariant. By Eq.~(\ref{eq:generate_pos}), the probability of each generated intermediate state will also be rotation invariance.
\label{app:zeromean}
\subsection{Details in diffusion models}
\paragraph{Diffusion Schedule.} As $\alpha_t$ and $\sigma_t$ are two parameters changing gradually along $t$-axis. Lots of noise schedules proposed, such as Linear \cite{ho2020denoising} and Cosine \cite{nichol2021improved}, here we choose a simple polynomial schedule
\begin{equation}
    \alpha_t^2 = (1 - (t/3)^3)^3,
\end{equation}
and variance-preserving process as 
\begin{equation}
\sigma_t^2 = 1 - \alpha_t^2. 
\end{equation}

\paragraph{Sampling Process.}
We give the detailed pseudo-codes of sampling algorithm of DiffBP in Algorithm 2. 
\begin{algorithm}[t]
  \caption{Sampling from DiffBP}
  \label{alg:sample_edm}
\begin{algorithmic}
\STATE {\bfseries Input:} Zero-centered protein $\{\bm{a}_j,\bm{x}_j\}_{j=1}^{M}$, pre-trained network $\varphi_{\omega}$ and graph denoiser $\phi_{\theta}$
\STATE Compute $[\bm{x}_{\mathrm{c}}^\mathrm{mol}, N^\mathrm{mol}] = \varphi_\omega(\{\bm{a}_{j}, \bm{x}_j\}_{j\in \mathcal{I}_{pro}})$
\STATE Subtract $\bm{x}_{\mathrm{c}}^\mathrm{mol}$ from $\{\bm{a}_j,\bm{x}_j\}_{j=1}^{M}$, and set $N := N^\mathrm{mol}$

\STATE Sample $\bm{x}_{i,T} \sim \mathcal{N}(\mathbf{0}, \bm{I})$
\FOR{$t$ in $T,\, T-1, \ldots, 1$ where $s = t - 1$}
\STATE Sample $\bm{\epsilon} \sim \mathcal{N}(\mathbf{0}, \bm{I})$
\STATE Subtract center of mass from $\vepsilon$
\STATE Compute $[\bm{\hat \epsilon}_{i,t},  \bm{\hat p}_{i,0}] = \phi_{\theta}(\{(\bm{a}_{j,t}, \bm{x}_{j,t})\}_{j=1}^{N+M}, t)$
\STATE Compute $\bm{x}_{i,s} = \frac{1}{\alpha_{t|s}} \bm{x}_{i,t} - \frac{\sigma_{t|s}^2}{\alpha_{t|s}\sigma_t} \bm{\hat \epsilon}_{i,t} + \sigma'_{t|s}  \vepsilon$
\STATE Sample uniformly $\mathcal{I}_{\mathrm{rcv}}$, $|\mathcal{I}_{\mathrm{rcv}}|/M = (T-t)/T$
\IF {$i\in \mathcal{I}_{\mathrm{rcv}}\cap\mathcal{I}_{\mathrm{mol}}$ \AND $a_{i,t}=K+1$}
\STATE Sample $a_{i,0} \sim \mathrm{Cat}(\bm{\hat p}_{i,0})$
\ENDIF

\ENDFOR
\STATE {\bfseries Output:} $\mathcal{M} = \{(\bm{a}_{i,0}, \bm{x}_{i,0})\}_{i=1}^{N}$
\end{algorithmic}
Note $\mathcal{I}_{\mathrm{rcv}}$ is the index set of atoms that need to be recovered to chemical types.
\end{algorithm}
\label{app:diffusion_detail}
\section{Experiments}
\label{app:exp}

\renewcommand\thefigure{B\arabic{figure}}
\renewcommand\thetable{B\arabic{table}}
\setcounter{table}{0}
\setcounter{figure}{0}

\subsection{Binding Affinity}
As in Pocket2Mol, the UFF refinement is not mentioned, so here we give the Binding Affinity of Pocket2Mol-without-UFF in Table.~\ref{tab:p2mapp}.

\begin{table}[h]
\centering
\caption{The Binding Affinity of Pocket2Mol-without-UFF.} \label{tab:p2mapp}\vspace{-0.5em}
\begin{tabular}{lrrr}
\toprule
       & \multicolumn{3}{c}{Pocket2Mol-without-UFF} \\
       \midrule
       &\multicolumn{1}{c}{Ratio}   & \multicolumn{1}{c}{MPBG}    & \multicolumn{1}{c}{$\Delta$Binding} \\
\cmidrule(lr){2-4}
Small  & 36.62\%      & 43.77\%      & 3.96\%       \\
Medium & 59.02\%      & 28.62\%      & 7.47\%       \\
Large  & 4.36\%       & 12.82\%      & 13.98\%      \\
\bottomrule
       &              & 33.47\%      & 6.46\%      \\
       \bottomrule
\end{tabular}
\end{table}
\subsection{Detailed drug properties}
Since there is a high correlation between Binding Scores and molecule sizes, here we list the other score functions of drug properties to figure out their preferences in Table.~\ref{tab:drugproperapp}.

It shows that score functions of QED, SA and LPSK have their preference for molecule sizes. QED and LPSK tend to give smaller molecules higher scores, while SA prefers larger ones.
\begin{table*}[t]
\caption{Detailed drug properties split into groups by molecule size.}\label{tab:drugproperapp}
\vspace{-0.5em}
\centering
\resizebox{0.8\linewidth}{!}{
\begin{tabular}{lrrrrrrrrrr}
\toprule
       & \multicolumn{5}{c}{3DSBDD}                                                                                                        & \multicolumn{5}{c}{Pocket2Mol}      \\
\midrule
       & \multicolumn{1}{c}{Ratio} & \multicolumn{1}{c}{QED} & \multicolumn{1}{c}{SA} & \multicolumn{1}{c}{SIM} & \multicolumn{1}{c}{LPSK} & \multicolumn{1}{c}{Ratio} & \multicolumn{1}{c}{QED} & \multicolumn{1}{c}{SA} & \multicolumn{1}{c}{SIM} & \multicolumn{1}{c}{LPSK} \\
       \cmidrule(lr){2-6} \cmidrule(lr){7-11}
Small  & 41.45\%                   & 0.4271                  & 4.2634                 & 0.3537                  & 0.7728                   & 36.62\%                   & 0.5479                  & 4.8287                 & 0.4176                  & 0.8929                   \\
Medium & 54.06\%                   & 0.3562                  & 4.9495                 & 0.3463                  & 0.6126                   & 59.02\%                   & 0.4923                  & 5.4200                 & 0.4463                  & 0.7731                   \\
Large  & 4.48\%                    & 0.2568                  & 5.8129                 & 0.3277                  & 0.3639                   & 4.36\%                    & 0.4443                  & 5.6048                 & 0.4033                  & 0.6921                   \\
\midrule
Overall  &                           & 0.3811                  & 4.7033                 & 0.3485                  & 0.6678                   &                           & 0.5106                  & 5.3125                 & 0.4104                  & 0.8134                   \\
\midrule
       & \multicolumn{5}{c}{GraphBP}                                                                                                       & \multicolumn{5}{c}{DiffBP}                                                                                                        \\
       \midrule

       & \multicolumn{1}{c}{Ratio} & \multicolumn{1}{c}{QED} & \multicolumn{1}{c}{SA} & \multicolumn{1}{c}{SIM} & \multicolumn{1}{c}{LPSK} & \multicolumn{1}{c}{Ratio} & \multicolumn{1}{c}{QED} & \multicolumn{1}{c}{SA} & \multicolumn{1}{c}{SIM} & \multicolumn{1}{c}{LPSK} \\
        \cmidrule(lr){2-6} \cmidrule(lr){7-11}
Small  & 27.72\%                   & 0.5092                  & 5.2193                 & 0.2757                  & 0.8893                   & 5.22\%                    & 0.5269                  & 5.6212                 & 0.3081                  & 0.8335                   \\
Medium & 32.03\%                   & 0.4596                  & 5.8747                 & 0.2628                  & 0.6931                   & 75.19\%                   & 0.4663                  & 6.0514                 & 0.3297                  & 0.7043                   \\
Large  & 37.97\%                   & 0.2492                  & 6.6990                  & 0.2899                  & 0.3360                    & 19.59\%                   & 0.3317                  & 6.0613                 & 0.3318                  & 0.6696                   \\
\midrule
Overall       &                           & 0.3830                  & 5.8610                 & 0.2707                  & 0.5961                   &                           & 0.4431                  & 6.0308                 & 0.3290                  & 0.7042                  \\
\bottomrule
\end{tabular}
}
\end{table*}

\subsection{Atom type analysis}
Expect the sub-structure, we give the ratio of different element types of atoms, for further comparison, as shown in Table.~\ref{tab:atomtypeapp}. Note that we only count the percentage of heavy atoms, in line with the generative models. 
\begin{table*}[h]
\centering
\caption{Atom type ratio of different methods.} \label{tab:atomtypeapp}
\vspace{-0.5em}
\resizebox{0.6\linewidth}{!}{
\begin{tabular}{lrrrrrr}
\toprule
   Atom type   & \multicolumn{1}{c}{Train} & \multicolumn{1}{c}{Test} & \multicolumn{1}{c}{3DSBDD} & \multicolumn{1}{c}{Pocket2Mol} & \multicolumn{1}{c}{GraphBP} & \multicolumn{1}{c}{DiffBP} \\
      \midrule
C     & 16.02                          & 14.03                         & 11.45                      & \textbf{12.29}                          & 20.57                       & \textbf{15.25}                      \\
N     & 2.82                           & 2.35                          & 1.91                       & \textbf{2.31}                           & 1.32                        & \textbf{2.12}                       \\
O     & 3.79                           & 4.98                          & 2.04                       & \textbf{2.92}                           & 1.86                        & \textbf{5.49}                       \\
F     & 0.31                           & 0.04                          & 0.03                       & 0.04                           & \textbf{0.16}                        & \textbf{0.09}                       \\
P     & 0.24                           & 0.39                          & 0.08                       & \textbf{0.11}                           & 0.06                        & 0.06                       \\
S     & 0.25                           & 0.22                          & \textbf{0.03}                       & 0.02                           & \textbf{0.11}                        & 1.45                       \\
Cl    & 0.15                           & 0.09                          & 0                          & 0                              & 0.07                        & 0.73                       \\
Br    & 0.05                           & 0                             & 0                          & 0                              & 0.04                        & 0                          \\
Se    & 0                              & 0                             & 0                          & 0                              & 0.02                        & 0.01                       \\
I     & 0.01                           & 0                             & 0                          & 0                              & 0.03                        & 0                          \\
Other & 0                              & 0                             & 0                          & 0                              & 0                           & 0                         \\
\bottomrule
\end{tabular}
}
\end{table*}
It shows that for the most common elements like `C',`O',`N', DiffBP and Pocket2Mol generate the average ratio that most closely resembles the true distribution. However, for the uncommon elements, such as `S' and `Cl', DiffBP generates more than reference.

\subsection{Implementation details}
\paragraph{Datasets.} The training set includes 10000 protein-ligand paired samples, of which 99981 are valid molecules. 100 protein pockets are used for the test, and for each sample, 100 molecules are generated in each trial.
\paragraph{Hyperparameters.}
In practice, we found that GVP is more numerically stable than EGNN, so we choose 12-layers GVP for the pre-generation model and 9-layers GVP for Graph denoiser. The hidden dimensions of vertices are fixed as 256 and 64 for invariant node attributes (atom types) and equivariant geometry positions, respectively.

In training, the batch size is set as 8, and the initial learning rate is set as $10^{-4}$. AdamW optimizer is used, together with the Exponential Moving Average training trick.

\paragraph{Platforms.}
We use a single NVIDIA A100(81920MiB) GPU for a trial. The codes are implemented in Python 3.9
mainly with Pytorch 1.12, and run on Ubuntu Linux.

%% file: main.bbl
\begin{thebibliography}{}\itemsep=-1pt

\end{thebibliography}


\begin{thebibliography}{10}\itemsep=-1pt

\bibitem{jing2021gvp1}
Anonymous.
\newblock Learning from protein structure with geometric vector perceptrons.
\newblock In {\em Submitted to International Conference on Learning
  Representations}, 2021.

\bibitem{jacob2021d3pm}
Jacob Austin, Daniel~D. Johnson, Jonathan Ho, Daniel Tarlow, and Rianne van~den
  Berg.
\newblock Structured denoising diffusion models in discrete state-spaces, 2021.

\bibitem{rosettafold1}
Minkyung Baek, Frank DiMaio, Ivan Anishchenko, Justas Dauparas, Sergey
  Ovchinnikov, Gyu~Rie Lee, Jue Wang, Qian Cong, Lisa~N. Kinch, R.~Dustin
  Schaeffer, Claudia Millán, Hahnbeom Park, Carson Adams, Caleb~R. Glassman,
  Andy DeGiovanni, Jose~H. Pereira, Andria~V. Rodrigues, Alberdina~A. van Dijk,
  Ana~C. Ebrecht, Diederik~J. Opperman, Theo Sagmeister, Christoph Buhlheller,
  Tea Pavkov-Keller, Manoj~K. Rathinaswamy, Udit Dalwadi, Calvin~K. Yip,
  John~E. Burke, K.~Christopher Garcia, Nick~V. Grishin, Paul~D. Adams,
  Randy~J. Read, and David Baker.
\newblock Accurate prediction of protein structures and interactions using a
  three-track neural network.
\newblock {\em Science}, 373(6557):871--876, 2021.

\bibitem{bengio2021gflownet}
Yoshua Bengio, Tristan Deleu, Edward~J. Hu, Salem Lahlou, Mo Tiwari, and
  Emmanuel Bengio.
\newblock {GFlowNet} foundations.
\newblock {\em CoRR}, abs/2111.09266, 2021.

\bibitem{bond2021unleashing}
Sam Bond-Taylor, Peter Hessey, Hiroshi Sasaki, Toby~P. Breckon, and Chris~G.
  Willcocks.
\newblock Unleashing transformers: Parallel token prediction with discrete
  absorbing diffusion for fast high-resolution image generation from
  vector-quantized codes.
\newblock In {\em European Conference on Computer Vision (ECCV)}, 2022.

\bibitem{cao2022survey}
Hanqun Cao, Cheng Tan, Zhangyang Gao, Guangyong Chen, Pheng-Ann Heng, and
  Stan~Z Li.
\newblock A survey on generative diffusion model.
\newblock {\em arXiv preprint arXiv:2209.02646}, 2022.

\bibitem{Dauparas2022pro}
J. Dauparas, I. Anishchenko, N. Bennett, H. Bai, R.~J. Ragotte, L.~F. Milles,
  B.~I.~M. Wicky, A. Courbet, R.~J. de Haas, N. Bethel, P.~J.~Y. Leung, T.~F.
  Huddy, S. Pellock, D. Tischer, F. Chan, B. Koepnick, H. Nguyen, A. Kang, B.
  Sankaran, A.~K. Bera, N.~P. King, and D. Baker.
\newblock Robust deep learning based protein sequence design using proteinmpnn.
\newblock {\em Science}, 2022.

\bibitem{dinh2016density}
Laurent Dinh, Jascha Sohl-Dickstein, and Samy Bengio.
\newblock Density estimation using real {NVP}.
\newblock {\em arXiv preprint arXiv:1605.08803}, 2016.

\bibitem{doersch2016tutorial}
Carl Doersch.
\newblock Tutorial on variational autoencoders.
\newblock {\em arXiv preprint arXiv:1606.05908}, 2016.

\bibitem{drews2000drug}
Jurgen Drews.
\newblock Drug discovery: a historical perspective.
\newblock {\em science}, 287(5460):1960--1964, 2000.

\bibitem{molsurvey}
Yuanqi Du, Tianfan Fu, Jimeng Sun, and Shengchao Liu.
\newblock {MolGenSurvey}: A systematic survey in machine learning models for
  molecule design, 2022.

\bibitem{AlphaFold-Multimer2021}
Richard Evans, Michael O{\textquoteright}Neill, Alexander Pritzel, Natasha
  Antropova, Andrew Senior, Tim Green, Augustin {\v{Z}}{\'\i}dek, Russ Bates,
  Sam Blackwell, Jason Yim, Olaf Ronneberger, Sebastian Bodenstein, Michal
  Zielinski, Alex Bridgland, Anna Potapenko, Andrew Cowie, Kathryn
  Tunyasuvunakool, Rishub Jain, Ellen Clancy, Pushmeet Kohli, John Jumper, and
  Demis Hassabis.
\newblock Protein complex prediction with alphafold-multimer.
\newblock {\em bioRxiv}, 2021.

\bibitem{francoeur2020three}
Paul~G Francoeur, Tomohide Masuda, Jocelyn Sunseri, Andrew Jia, Richard~B
  Iovanisci, Ian Snyder, and David~R Koes.
\newblock Three-dimensional convolutional neural networks and a cross-docked
  data set for structure-based drug design.
\newblock {\em Journal of Chemical Information and Modeling}, 60(9):4200--4215,
  2020.

\bibitem{octavian2021equidock}
Octavian-Eugen Ganea, Xinyuan Huang, Charlotte Bunne, Yatao Bian, Regina
  Barzilay, Tommi Jaakkola, and Andreas Krause.
\newblock Independent se(3)-equivariant models for end-to-end rigid protein
  docking.
\newblock 2021.

\bibitem{gao2022alphadesign}
Zhangyang Gao, Cheng Tan, Stan Li, et~al.
\newblock Alphadesign: A graph protein design method and benchmark on
  alphafolddb.
\newblock {\em arXiv preprint arXiv:2202.01079}, 2022.

\bibitem{mpnn}
Justin Gilmer, Samuel~S. Schoenholz, Patrick~F. Riley, Oriol Vinyals, and
  George~E. Dahl.
\newblock Neural message passing for quantum chemistry, 2017.

\bibitem{goodfellow2014generative}
Ian Goodfellow, Jean Pouget-Abadie, Mehdi Mirza, Bing Xu, David Warde-Farley,
  Sherjil Ozair, Aaron Courville, and Yoshua Bengio.
\newblock Generative adversarial nets.
\newblock In {\em NIPS}, pages 2672--2680, 2014.

\bibitem{ho2020denoising}
Jonathan Ho, Ajay Jain, and Pieter Abbeel.
\newblock Denoising diffusion probabilistic models.
\newblock {\em arXiv preprint arXiv:2006.11239}, 2020.

\bibitem{emiel2021edm}
Emiel Hoogeboom, Victor~Garcia Satorras, Clément Vignac, and Max Welling.
\newblock Equivariant diffusion for molecule generation in 3d, 2022.

\bibitem{rosettafold2}
Ian~R. Humphreys, Jimin Pei, Minkyung Baek, Aditya Krishnakumar, Ivan
  Anishchenko, Sergey Ovchinnikov, Jing Zhang, Travis~J. Ness, Sudeep Banjade,
  Saket~R. Bagde, Viktoriya~G. Stancheva, Xiao-Han Li, Kaixian Liu, Zhi Zheng,
  Daniel~J. Barrero, Upasana Roy, Jochen Kuper, Israel~S. Fernández, Barnabas
  Szakal, Dana Branzei, Josep Rizo, Caroline Kisker, Eric~C. Greene, Sue
  Biggins, Scott Keeney, Elizabeth~A. Miller, J.~Christopher Fromme, Tamara~L.
  Hendrickson, Qian Cong, and David Baker.
\newblock Computed structures of core eukaryotic protein complexes.
\newblock {\em Science}, 374(6573):eabm4805, 2021.

\bibitem{pmlr-v119-jin20a}
Wengong Jin, Dr.Regina Barzilay, and Tommi Jaakkola.
\newblock Hierarchical generation of molecular graphs using structural motifs.
\newblock In Hal~Daumé III and Aarti Singh, editors, {\em Proceedings of the
  37th International Conference on Machine Learning}, volume 119 of {\em
  Proceedings of Machine Learning Research}, pages 4839--4848. PMLR, 13--18 Jul
  2020.

\bibitem{jing2021gvp2}
Bowen Jing, Stephan Eismann, Pratham~N. Soni, and Ron~O. Dror.
\newblock Equivariant graph neural networks for 3d macromolecular structure,
  2021.

\bibitem{AlphaFold2021}
John Jumper, Richard Evans, Alexander Pritzel, Tim Green, Michael Figurnov,
  Olaf Ronneberger, Kathryn Tunyasuvunakool, Russ Bates, Augustin
  {\v{Z}}{\'\i}dek, Anna Potapenko, Alex Bridgland, Clemens Meyer, Simon A~A
  Kohl, Andrew~J Ballard, Andrew Cowie, Bernardino Romera-Paredes, Stanislav
  Nikolov, Rishub Jain, Jonas Adler, Trevor Back, Stig Petersen, David Reiman,
  Ellen Clancy, Michal Zielinski, Martin Steinegger, Michalina Pacholska, Tamas
  Berghammer, Sebastian Bodenstein, David Silver, Oriol Vinyals, Andrew~W
  Senior, Koray Kavukcuoglu, Pushmeet Kohli, and Demis Hassabis.
\newblock Highly accurate protein structure prediction with {AlphaFold}.
\newblock {\em Nature}, 596(7873):583--589, 2021.

\bibitem{kingma2018glow}
Durk~P Kingma and Prafulla Dhariwal.
\newblock Glow: Generative flow with invertible 1x1 convolutions.
\newblock {\em Advances in Neural Information Processing Systems},
  31:10215--10224, 2018.

\bibitem{Dieerik2021vdm}
Diederik~P. Kingma, Tim Salimans, Ben Poole, and Jonathan Ho.
\newblock Variational diffusion models, 2021.

\bibitem{gcn}
Thomas~N. Kipf and Max Welling.
\newblock Semi-supervised classification with graph convolutional networks,
  2016.

\bibitem{kuntz1992structure}
Irwin~D Kuntz.
\newblock Structure-based strategies for drug design and discovery.
\newblock {\em Science}, 257(5073):1078--1082, 1992.

\bibitem{kohler2020ef}
Jonas Köhler, Leon Klein, and Frank Noé.
\newblock Equivariant flows: Exact likelihood generative learning for symmetric
  densities, 2020.

\bibitem{landrum2006rdkit}
Greg Landrum et~al.
\newblock {RDKit}: Open-source cheminformatics.
\newblock 2006.

\bibitem{liu2022graphbp}
Meng Liu, Youzhi Luo, Kanji Uchino, Koji Maruhashi, and Shuiwang Ji.
\newblock Generating 3d molecules for target protein binding.
\newblock In {\em International Conference on Machine Learning}, 2022.

\bibitem{liu2021dig}
Meng Liu, Youzhi Luo, Limei Wang, Yaochen Xie, Hao Yuan, Shurui Gui, Haiyang
  Yu, Zhao Xu, Jingtun Zhang, Yi Liu, et~al.
\newblock {DIG}: A turnkey library for diving into graph deep learning
  research.
\newblock {\em Journal of Machine Learning Research}, 22(240):1--9, 2021.

\bibitem{liu2021graphebm}
Meng Liu, Keqiang Yan, Bora Oztekin, and Shuiwang Ji.
\newblock {GraphEBM}: Molecular graph generation with energy-based models.
\newblock {\em arXiv preprint arXiv:2102.00546}, 2021.

\bibitem{liu2021spherical}
Yi Liu, Limei Wang, Meng Liu, Yuchao Lin, Xuan Zhang, Bora Oztekin, and
  Shuiwang Ji.
\newblock Spherical message passing for 3d molecular graphs.
\newblock In {\em International Conference on Learning Representations}, 2021.

\bibitem{luo20213d}
Shitong Luo, Jiaqi Guan, Jianzhu Ma, and Jian Peng.
\newblock A {3D} generative model for structure-based drug design.
\newblock In {\em Thirty-Fifth Conference on Neural Information Processing
  Systems}, 2021.

\bibitem{luo2021predicting}
Shitong Luo, Chence Shi, Minkai Xu, and Jian Tang.
\newblock Predicting molecular conformation via dynamic graph score matching.
\newblock {\em Advances in Neural Information Processing Systems}, 34, 2021.

\bibitem{luo2021autoregressive}
Youzhi Luo and Shuiwang Ji.
\newblock An autoregressive flow model for 3d molecular geometry generation
  from scratch.
\newblock In {\em International Conference on Learning Representations}, 2021.

\bibitem{masuda2020generating}
Tomohide Masuda, Matthew Ragoza, and David~Ryan Koes.
\newblock Generating 3d molecular structures conditional on a receptor binding
  site with deep generative models.
\newblock {\em arXiv preprint arXiv:2010.14442}, 2020.

\bibitem{mcnutt2021gnina}
Andrew~T McNutt, Paul Francoeur, Rishal Aggarwal, Tomohide Masuda, Rocco Meli,
  Matthew Ragoza, Jocelyn Sunseri, and David~Ryan Koes.
\newblock Gnina 1.0: molecular docking with deep learning.
\newblock {\em Journal of cheminformatics}, 13(1):1--20, 2021.

\bibitem{nichol2021improved}
Alexander~Quinn Nichol and Prafulla Dhariwal.
\newblock Improved denoising diffusion probabilistic models, 2021.

\bibitem{nielsen2020survae}
Didrik Nielsen, Priyank Jaini, Emiel Hoogeboom, Ole Winther, and Max Welling.
\newblock Survae flows: Surjections to bridge the gap between vaes and flows.
\newblock {\em Advances in Neural Information Processing Systems},
  33:12685--12696, 2020.

\bibitem{o2011open}
Noel~M O'Boyle, Michael Banck, Craig~A James, Chris Morley, Tim Vandermeersch,
  and Geoffrey~R Hutchison.
\newblock Open babel: An open chemical toolbox.
\newblock {\em Journal of cheminformatics}, 3(1):1--14, 2011.

\bibitem{peng2022pocket2mol}
Xingang Peng, Shitong Luo, Jiaqi Guan, Qi Xie, Jian Peng, and Jianzhu Ma.
\newblock Pocket2mol: Efficient molecular sampling based on 3d protein pockets.
\newblock In {\em International Conference on Machine Learning}, 2022.

\bibitem{polykovskiy2020molecular}
Daniil Polykovskiy, Alexander Zhebrak, Benjamin Sanchez-Lengeling, Sergey
  Golovanov, Oktai Tatanov, Stanislav Belyaev, Rauf Kurbanov, Aleksey
  Artamonov, Vladimir Aladinskiy, Mark Veselov, et~al.
\newblock Molecular sets ({MOSES}): a benchmarking platform for molecular
  generation models.
\newblock {\em Frontiers in pharmacology}, 2020.

\bibitem{ragoza2017protein}
Matthew Ragoza, Joshua Hochuli, Elisa Idrobo, Jocelyn Sunseri, and David~Ryan
  Koes.
\newblock Protein--ligand scoring with convolutional neural networks.
\newblock {\em Journal of chemical information and modeling}, 57(4):942--957,
  2017.

\bibitem{rappe1992uff}
Anthony~K Rapp{\'e}, Carla~J Casewit, KS Colwell, William~A Goddard~III, and
  W~Mason Skiff.
\newblock {UFF}, a full periodic table force field for molecular mechanics and
  molecular dynamics simulations.
\newblock {\em Journal of the American chemical society}, 114(25):10024--10035,
  1992.

\bibitem{Victor2021enf}
Victor~Garcia Satorras, Emiel Hoogeboom, Fabian~B. Fuchs, Ingmar Posner, and
  Max Welling.
\newblock E(n) equivariant normalizing flows, 2021.

\bibitem{victor2022egnn}
Victor~Garcia Satorras, Emiel Hoogeboom, and Max Welling.
\newblock E(n) equivariant graph neural networks, 2021.

\bibitem{shi2021learning}
Chence Shi, Shitong Luo, Minkai Xu, and Jian Tang.
\newblock Learning gradient fields for molecular conformation generation.
\newblock {\em Proceedings of the 38th International Conference on Machine
  Learning, {ICML}}, 139:9558--9568, 2021.

\bibitem{shi2019graphaf}
Chence Shi, Minkai Xu, Zhaocheng Zhu, Weinan Zhang, Ming Zhang, and Jian Tang.
\newblock {GraphAF}: a flow-based autoregressive model for molecular graph
  generation.
\newblock In {\em ICLR}, 2020.

\bibitem{sohl2015deep}
Jascha Sohl-Dickstein, Eric Weiss, Niru Maheswaranathan, and Surya Ganguli.
\newblock Deep unsupervised learning using nonequilibrium thermodynamics.
\newblock In {\em International Conference on Machine Learning}, pages
  2256--2265. PMLR, 2015.

\bibitem{song2019generative}
Yang Song and Stefano Ermon.
\newblock Generative modeling by estimating gradients of the data distribution.
\newblock {\em Advances in Neural Information Processing Systems}, 32, 2019.

\bibitem{song2020score}
Yang Song, Jascha Sohl-Dickstein, Diederik~P Kingma, Abhishek Kumar, Stefano
  Ermon, and Ben Poole.
\newblock Score-based generative modeling through stochastic differential
  equations.
\newblock {\em arXiv preprint arXiv:2011.13456}, 2020.

\bibitem{YaotingSun2022ArtificialID}
Yaoting Sun, Sathiyamoorthy Selvarajan, Zelin Zang, Wei Liu, Yi Zhu, Hao Zhang,
  Wanyuan Chen, Hao Chen, Lu Li, Xue Cai, Huanhuan Gao, Zhicheng Wu, Yongfu
  Zhao, Lirong Chen, Xiaodong Teng, Sangeeta Mantoo, Tony Kiat, Hon Lim,
  Bhuvaneswari Hariraman, Serene Yeow, Syed Muhammad, Fahmy Alkaff, Sze~Sing
  Lee, Guan Ruan, Qiushi Zhang, Tiansheng Zhu, Yifan Hu, Zhen Dong, Weigang Ge,
  Qi Xiao, Weibin Wang, Guangzhi Wang, Junhong Xiao, Yi He, Zhihong Wang, Wei
  Sun, Yuan Qin, Jiang Zhu, Xu Zheng, Linyan Wang, Xi Zheng, Kailun Xu,
  Yingkuan Shao, Shu Zheng, Kexin Liu, Ruedi Aebersold, Haixia Guan, Xiaohong
  Wu, Dingcun Luo, Wen Tian, Stan~Ziqing Li, Oi~Lian Kon,
  Narayanan~Gopalakrishna Iyer, and Tiannan Guo.
\newblock Artificial intelligence defines protein-based classification of
  thyroid nodules.
\newblock 2022.

\bibitem{proteinsurface}
Freyr Sverrisson, Jean Feydy, Bruno~E. Correia, and Michael~M. Bronstein.
\newblock Fast end-to-end learning on protein surfaces.
\newblock In {\em 2021 IEEE/CVF Conference on Computer Vision and Pattern
  Recognition (CVPR)}, pages 15267--15276, 2021.

\bibitem{tan2022target}
Cheng Tan, Zhangyang Gao, and Stan~Z Li.
\newblock Target-aware molecular graph generation.
\newblock {\em arXiv preprint arXiv:2202.04829}, 2022.

\bibitem{townshend2021atom3d}
Raphael John~Lamarre Townshend, Martin V{\"o}gele, Patricia~Adriana Suriana,
  Alexander Derry, Alexander Powers, Yianni Laloudakis, Sidhika Balachandar,
  Bowen Jing, Brandon~M Anderson, Stephan Eismann, et~al.
\newblock Atom3d: Tasks on molecules in three dimensions.
\newblock In {\em Thirty-fifth Conference on Neural Information Processing
  Systems Datasets and Benchmarks Track (Round 1)}, 2021.

\bibitem{proteinsurface2}
Vishwesh Venkatraman, Yifeng Yang, Lee Sael, and Daisuke Kihara.
\newblock Protein-protein docking using region-based 3d zernike descriptors.
\newblock {\em BMC bioinformatics}, 10:407, 12 2009.

\bibitem{wu2021selfsuper}
Lirong Wu, Haitao Lin, Cheng Tan, Zhangyang Gao, and Stan~Z. Li.
\newblock Self-supervised learning on graphs: Contrastive, generative,or
  predictive.
\newblock {\em IEEE Transactions on Knowledge and Data Engineering}, pages
  1--1, 2021.

\bibitem{xu2021learning}
Minkai Xu, Shitong Luo, Yoshua Bengio, Jian Peng, and Jian Tang.
\newblock Learning neural generative dynamics for molecular conformation
  generation.
\newblock {\em International Conference on Learning Representations, ({ICLR})},
  2021.

\bibitem{xu2021end}
Minkai Xu, Wujie Wang, Shitong Luo, Chence Shi, Yoshua Bengio, Rafael
  Gomez-Bombarelli, and Jian Tang.
\newblock An end-to-end framework for molecular conformation generation via
  bilevel programming.
\newblock 2021.

\bibitem{xu2021geodiff}
Minkai Xu, Lantao Yu, Yang Song, Chence Shi, Stefano Ermon, and Jian Tang.
\newblock {GeoDiff}: A geometric diffusion model for molecular conformation
  generation.
\newblock In {\em International Conference on Learning Representations}, 2021.

\bibitem{xu2021molecule3d}
Zhao Xu, Youzhi Luo, Xuan Zhang, Xinyi Xu, Yaochen Xie, Meng Liu, Kaleb
  Dickerson, Cheng Deng, Maho Nakata, and Shuiwang Ji.
\newblock {Molecule3D}: A benchmark for predicting 3d geometries from molecular
  graphs.
\newblock {\em arXiv preprint arXiv:2110.01717}, 2021.

\bibitem{yang2017chemts}
Xiufeng Yang, Jinzhe Zhang, Kazuki Yoshizoe, Kei Terayama, and Koji Tsuda.
\newblock {ChemTS}: an efficient python library for de novo molecular
  generation.
\newblock {\em Science and technology of advanced materials}, 18(1):972--976,
  2017.

\bibitem{zang2020moflow}
Chengxi Zang and Fei Wang.
\newblock {MoFlow}: an invertible flow model for generating molecular graphs.
\newblock In {\em ACM SIGKDD}, pages 617--626, 2020.

\bibitem{Zeng2022deep}
Xiangxiang Zeng, Fei Wang, Yuan Luo, Seung gu Kang, Jian Tang, Felice~C.
  Lightstone, Evandro~F. Fang, Wendy Cornell, Ruth Nussinov, and Feixiong
  Cheng.
\newblock Deep generative molecular design reshapes drug discovery.
\newblock {\em Cell Reports Medicine}, 2022.

\end{thebibliography}
